# Subwavelength gratings for OVDs
# From local interactions to using light-transport

Guillaume Basset[a], Giorgio Quaranta[a], Fabian Lütolf[a], Laurent Davoine and Marc Schnieper[a]

[a] CSEM SA, Tramstrasse 99, Muttenz, Switzerland

## ABSTRACT

In the past 30 years, subwavelength gratings have been developed and produced as highly secured DOVIDs, allowing new distinct optical effects. In particular, subwalength gratings coated with a high refractive index dielectric are well-known and mass-produced to secure documents, such as the DIDs™ (diffractive Identification Devices). These submicronic gratings are usually called Zero Order Structures or Filters, as well as diffractive microstructures designed for Zero-Order read-out. Their structure and optical behavior, when implemented as Diffractive Optical Variable Image Devices (DOVIDs), have been already presented and explained[1].

Similar structures are also called Resonant Waveguide Gratings (RWG) or Resonant Leaky Mode Waveguides, often when optimized for different purposes. This behavior was not reported in detail and we propose here to explain it briefly. A detail study of the behavior of structures such as DIDs™ reveals that, when viewed aligned to the gratings (in conical incidence) their behavior is indeed a pure Zero-order diffractive effect. However, more complex phenomena are observed when viewed across the gratings (in collinear incidence). The results of rigorous and time-resolved optical simulations presented here can quantify and illustrate such optical coupling and propagation in the RWG. The excited modes of the RWG are playing a significant role in the appearance of DIDs™ in collinear incidence. These results have practical impacts on the design and engineering of such security elements.

From the learning of such non-local light interactions in DIDs™, new optically variable effects can be designed and produced by using light transport in RWG. As an example, multi-frequency RWGs can be designed to couple light for different colors and polarizations and to outcouple them away from the specular configuration typical to the Zero Order Devices. Example of such structures and the new optical effects that can be produced are presented and discussed.

As a further step and with a total change of the usual design of DOVIDs, new structures allow to couple light in the substrates or in selected layers of the substrates of security labels or documents. These DOVIDs enable selective light transport at a macroscopic scale using total internal reflection in the volume of the substrates. Example of suitable light-coupling structures compatible with current mass-production methods are presented. Optical design rules for such highly multimode waveguides are explained. Advantages of light transport in the polymer bulk of security documents are discussed with the case of tamper seals that are widely used in Brand Protection.

## 1. CURRENT SUBWAVELENGTHS GRATINGS AS DOVIDs: THE DIDs[3]

### 1.1 A Brief History Of The DIDs

The creation of the first subwavelength DOVID in the early eighties followed the request of the Federal Reserve to use new optical technologies to protect the US Dollars. At the beginning of the 1980s, the fraction of counterfeited the security of US Dollars was challenged by the advance of new printing techniques, especially photocopier and color photocopier were getting wider access, lower operating costs and better fidelity. In order to reduce the counterfeited fraction of US Dollars, Paul Volcker, then chairman of the Federal Reserve[4] sent a letter to leading American technology companies to develop new security devices. RCA Corporation, actively searching to replace magnetic tape by optical tape for data storage forwarded the request to their optical laboratory in Zurich, Switzerland.

In 1981, the first subwavelength gratings designed to be read-out in Zero-order were prototyped in the RCA Zurich laboratory, as shown in Fig. 1.

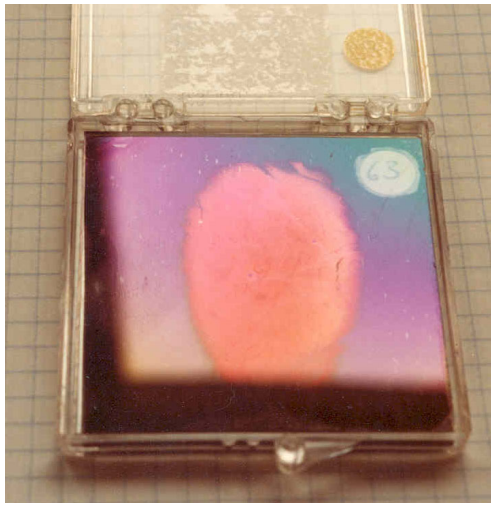

Figure 1: Pictures of the first demonstrator of Zero-order filter, using a subwavelength grating and a resonant waveguide, fabricated in 1981 by K. Knop. On the left, the specular reflected color in conical incidence, and in collinear incidence on the right.

The related patent demand was filed in February 1981 and granted in 1984[2]. An extract is shown in Fig. 2.

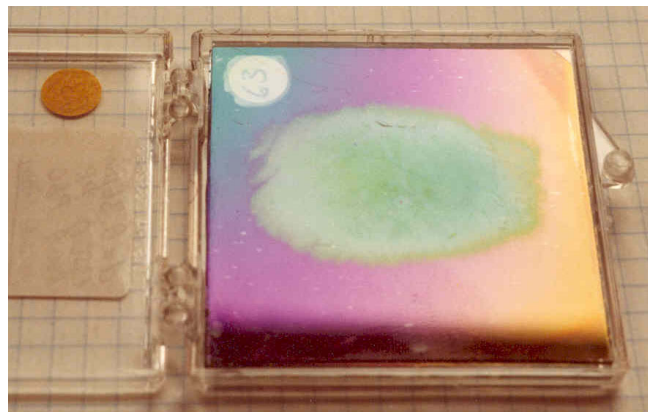

Figure 2: Extract from the US patent 4426130 filled in February 1981 and granted in January 1984, disclosing for the first time a Zero-order filter[2].

Following the takeover by General Electric and the break-up of the RCA group in the late eighties, the optical laboratory and know-how was transferred to the Paul Scherer Institute and finally to CSEM SA. CSEM partnered with

Hologram Industries to market this technology in 2003 as a successful DOVIDs, under the brand name DID™, for Diffraction Identification Devices, recognized as a third generation DOVIDs.

Nowadays, the DIDs™ are securing passport laminates of various countries and were as well introduced to banknotes and ID cards. New developments based on Zero-Order read-out schemes have been developed in the past ten years, including multiple resonant waveguides gratings[5], two dimensional resonant gratings[6] and recently plasmonic resonance based DOVIDs[7,8].

### 1.2 DIDs™, "Zero-Order Devices"

Zero-order devices, in opposition to first or higher diffraction order devices, can be visually controlled in any light condition, including fully diffused ambient lighting. This unique property is enabled by the specific optical read-out of these devices made on the Zero-order which is diffractively filtered by the subwavelength structures. In details, the incident light is splitted, between the transmitted Zero-order, the reflected Zero-order and, depending of the incidence angle, the transmitted and reflected positive first order and diffractive first order. M.T. Gale, K. Knop and R. Morf illustrated this[1] with a simple schematic reported in Fig. 3.

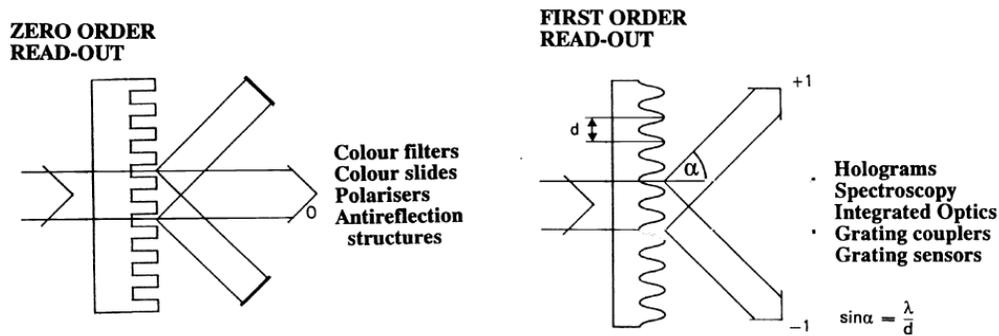

Figure 3: Zero-order read-out scheme as reported by M.T. Gale, K. Knop and R. Morf[1].

In reality, most third generation DOVIDs such as the DIDs™ are read-out using specular, or pseudo-specular reflection when a chaotic modulation of the subwavelength structures opens the viewing angle[7], as illustrated in Fig. 4.

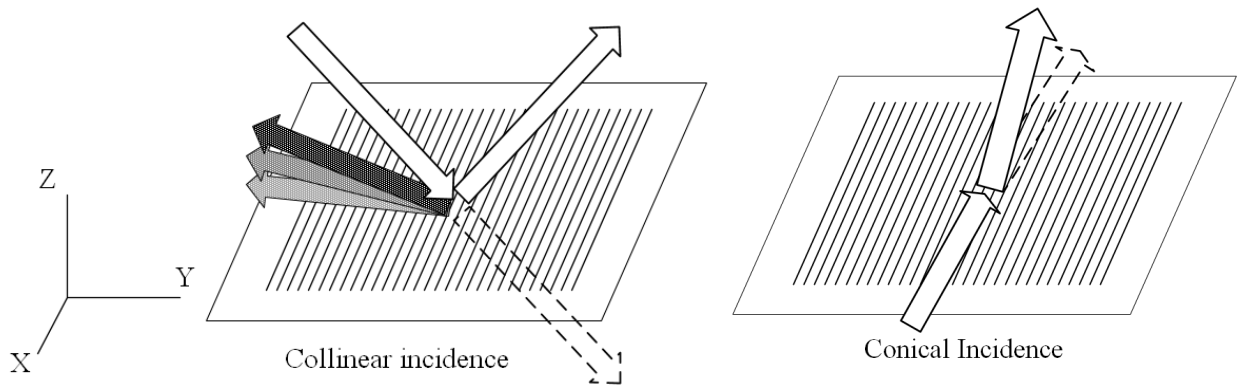

Figure 4: Schematic view of specular or quasi-specular reflection read-out of Zero-order DOVIDs. On the left, collinear incidence is illustrated, while conical incidence is represented on the right. The dotted arrow represents the Zero-order transmitted beam. Possible first order diffracted light, spectrally dispersed, is represented in grey-arrow, with the example of the reflected diffracted beam in the collinear incidence case.

The observed Zero-order reflected beam spectrum is composed of the incident beam spectrum subtracted with the Zero-order transmitted beam and the eventual first order diffracted beams (reflected and transmitted). Light absorption and scattering are generally very low in the state-of the art devices. The nanostructure diffractive filtering is usually influenced strongly by the polarization of the light. As a result, specular reflected is partially polarized, in some cases fully linearly polarized, which enables an efficient level two control using standard polarizers or polarized light source.

In this read-out configuration, the observation angle fully determines the appearance of the DOVIDs, on the opposite to second generation DOVIDs for which the illumination condition is critical. This enables visual authentication control in any light condition, including under multiple light sources and in fully diffuse illuminations. In addition to the mastering complexity, to the need for high fidelity replications and to the accurate coating industrial capabilities, this visual effect proved highly attractive for high security. Twelve years after its introduction to market, no counterfeit or even imitations have been reported.

### 1.3 DIDs™ In Collinear Incidence: Light Coupling And Delocalized Effect

As described above, third generation DOVIDs such as DIDs are read-out in Zero-order. In a first approximation, the subwavelength nanostructures, whether having a sinusoidal shape such as reported in 1981[2], a binary shape reported in 1990[1] or another profile, are filtering light through Zero-order diffractive processes. As illustrated in figure 4, at some observation angles, first order diffraction will take part in the spectral filtering process.

However, in a more detailed investigation, these structures, sometimes called Resonant Waveguide Gratings (RWG) or Resonant Leaky Mode Waveguides, do not filter all incident light in a purely local diffractive process. Various published works detailed the optical guided mode resonance of very similar structures[9,10,11,12]. Different devices have been reported, including linewidth bandpass, wideband polarization independent reflector and wideband anti-reflector filters as well as polarizers. These devices cannot be directly used in optical security because of the angular dependency of these devices, their too narrow bandwidth for visual authentication or the layer stack required. On the opposite, the optical-guided mode resonances of third generation DOVIDs such as DIDs™ were not reported so far according to our knowledge.

We propose here a very short analysis of such a structure using a common numerical analysis tool, Finite Difference Time Domain (FDTD), to extract time-dependent information (see Fig. 5). A typical Zero-order device is approximated with a binary shape waveguide-grating whose high refractive index (HRI) waveguide-grating is made of Zinc-Sulfide while the surrounding media is a typical polymer or lacquer used in DOVIDs technical layer.

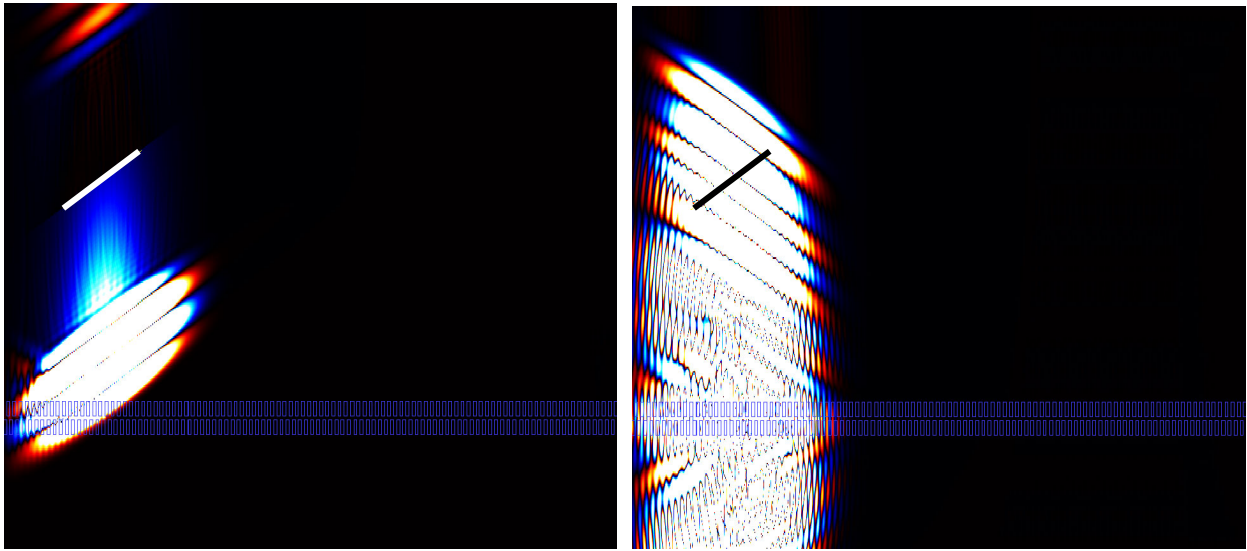

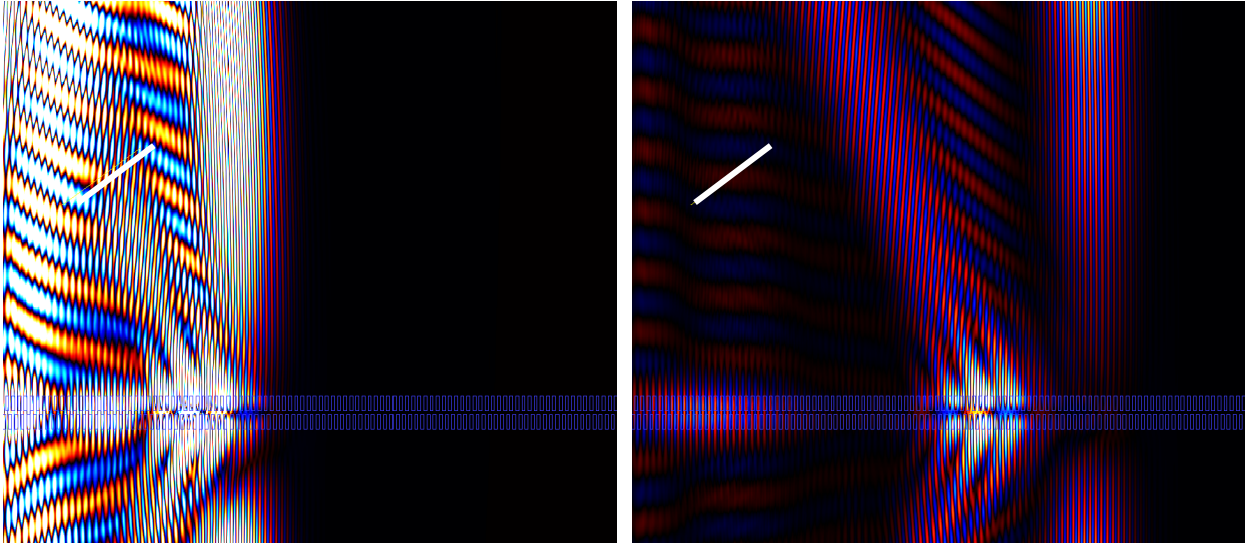

Figure 5: The four pictures illustrate a cross-section of a third generation DIDs when excited by a white Gaussian light pulse at four consecutive instants. 2D views display the electric field spatial distribution, blue and red pixels display its polarity while the brightness represents the E-field intensity. The HRI dielectric domains are schematized in dark blue rectangles. A white light quasi-planar wavefront pulse with a Gaussian spectral distribution is generated by a finite excitor (represented in white, top left side) at a time $T_0$. The excitor light pulse is polarized with a transverse electric orientation to match the resonant coupling condition of the structures.

At a first instant $T_1$ (top left), the quasi planar finite wavefronts can be seen propagating towards the subwavelength structures. At a second instant $T_2$ (top right), the wavefront diffractive filtering between the Zero-order transmission and reflection can be observed. After a short delay in $T_3$ (bottom left), the incident and reflected wavefronts moving away from the nanostructure, the formation of a resonant coupled mode can be observed. After a further delay $T_4$ (bottom right), the leaky guided mode can be seen after a few microns of propagation. The overall process takes place in less than 200fs after the initial excitation. The mode-order coupled has a Gaussian distribution around its central wavelength. By analyzing the propagating mode intensity in different locations, or at different instants, it is possible to compute its propagation length.

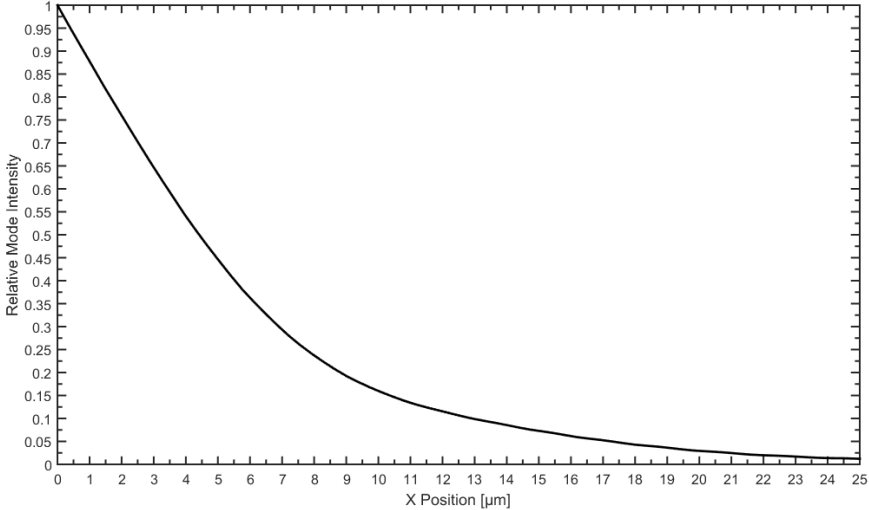

Figure 6: Leaky Mode Intensity over propagation distance with an exponential decay.

In this specific example, the leaky mode has a propagation decay length of 5.58μm, also named coupling length, as shown in Fig. 6. Experimental analyses of guided mode resonances at the micron scale in real devices are difficult, while matching closely numerical analysis when reported. This excited leaky guided modes, when encountering other structures, will be diffracted or scattered out. This must be taken into account while performing micrometer-level patterning of such structures. A short analytical analysis is presented in the next section with a proposal to exploit these modes.

## 2. LIGHT TRANSPORT IN RESONANT WAVEGUIDE-GRATINGS: NEW EFFECTS

### 2.1 Generalities On Resonant Waveguide Gratings

Resonant waveguide gratings consist of a combination of a subwavelength grating and a thin film waveguide. A resonance occurs when the incident light is diffracted by the grating and matches a mode of the waveguide. As most of the spectrum does not couple into the waveguide, strong spectral changes are observed in reflection and transmission. The existence of such resonances has been first discovered by Wood in 1902[13]. They belong to one type of the anomalous diffraction phenomena in grating structures. The term anomaly implies a rapid variation in the external observable diffracted orders with respect to physical parameters such as the angle and/or the wavelength and/or the polarization of the incident wave. At that time abrupt change of reflection could not be explained. Intensive research have been done to explain the anomalous phenomenon.[14-16]

Hessel and Oliner pointed out that they are basically two types of grating anomalies.[15] One is called the Rayleigh type, which is the classical Woods anomaly, and another is called the resonance type. The existence of Rayleigh anomaly is owing to the energy of higher diffracted orders that is transferred to an evanescent wave as a result of energy redistribution in remaining propagated diffraction orders. The appearance of the resonance anomaly in diffraction grating is due to the coupling process of externally incident wave to a surface guided wave which is supported by the structure.

Several researchers have studied these resonance effects in the past. Hessel and Oliner constructed the first model to describe the resonance type anomaly[15]. They modeled the planar grating as a periodic reactance surface and calculated results for reflection gratings. The major difficulty was the mathematical description of grating structure with a period smaller than the wavelength. In the case of shallow grating and a planar waveguide, accurate results can be obtained by using the Fourier-Rayleigh approximation. The method falls in the case of deep grating grooves. Several authors investigated the reflection from weakly corrugated waveguides. The convergence problems of deep grating grooves could be relaxed by using the rigorous Fourier Modal Method (FMM) or the Rigorous Coupled Wave Analysis (RCWA). With these new tools, many devices were derived. One of the main applications was the design of filters with very narrow spectral linewidths in reflection and transmission.

Rosenblatt[20] and Sharon[21] made a systematic analysis of resonant grating waveguide structures. They reviewed the basic principles of resonant gratings, and explained that the efficient transfer of wave energy between forward and backward propagations at resonance is due to the relative π phase-shift between the incident and the diffracted waves, which results in destructive and constructive interference of forward and backward propagating waves, respectively. They used both a geometrical ray approach and coupled-wave models to analyze the resonance behavior in grating-waveguide structures.

### 2.2 A Short Theory Of Thin Waveguide-Gratings

When a diffraction grating and a slab waveguide are brought into proximity and the angle of a diffracted mode matches the angle of a guided mode, the phenomenon of the so called Guided-Mode resonance happens. Fig. 7 shows a scheme of a resonance waveguide grating, made with a slab waveguide and a diffraction grating. The waveguide has a defined thickness t, refractive index $n_{core}$; the diffraction grating has period $\Lambda$ and depth $d$, and in the general case, due to the fabrication process, it is present in both side of the waveguide.

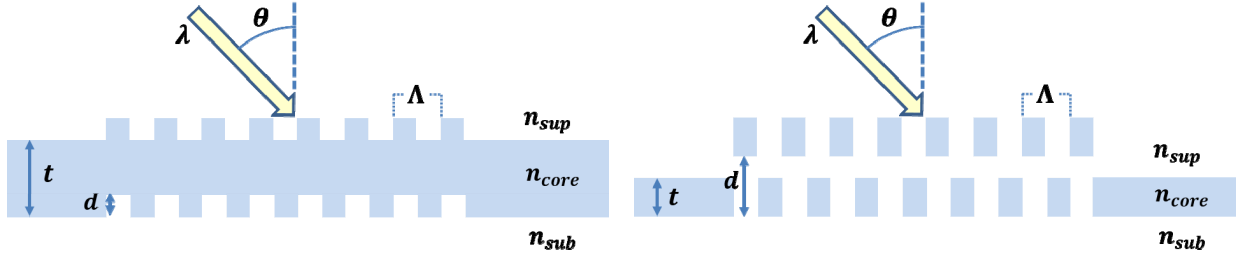

Figure 7: Schemes of resonant waveguide gratings with a binary profiles. Superstrate and substrate are defined according to the light incidence direction. The left configuration has the waveguide thickness higher than the grating depth, whereas the right configuration has the grating depth higher than the waveguide thickness. Other grating profiles such as sinusoidal, rounded binary grating or triangular shape can be evaluated similarly.

To illustrate this coupling behavior and its strong angular, wavelength and polarization dependence (so-called anomaly), we illustrate it here by numerical tools. Using FDTD, it is clear that mode coupling is in this case highly wavelength dependent. In Fig. 8 the electric field intensity is plotted, for a quasi-plain wave impinging with normal incidence on a resonant waveguide grating. The blue domain represents a HRI dielectric while the surrounding media is a typical transparent polymer or lacquer. When the structure is not at resonance (left image), it exhibits the background response of a multilayer device. In the near-field, the transmitted and reflected wavefront are distorted but are quasi-planar in the far-field. On the other hand, when the resonance condition is met, a large fraction of the applied wave is coupled into a guided mode (right image) that can propagate in the thin waveguide. This waveguide is not corrugated on the left and right side to allow propagation without leakage over large distances.

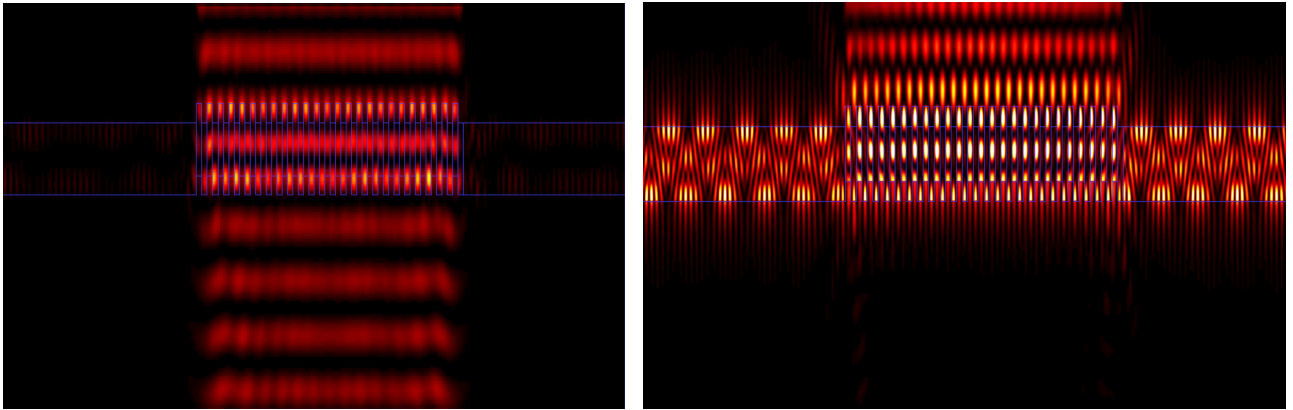

Figure 8: FDTD simulation of the resonances of waveguide grating. The wavelength is out of resonance regime on the left simulation, at a first wavelength and within resonance regime on the right at a second wavelength, impeding with a normal incidence and transverse electric polarization. In this example, at resonance the incident light-beam is diffracted at a first positive and negative orders, and the coupled mode is the third transverse electric mode of the waveguide.

At resonance, almost 100% switching of light energy between reflected and transmitted or diffracted and transmitted wave occurs over small parameter ranges[22]. The resonance occurs when the diffracted orders inside the waveguide structure are phase-matched to the guided mode supported by the waveguide structure. The effective mode index of refraction of the waveguide can be expressed as $n_{\text{eff}} = n_{\text{sup}} \sin\theta - m\frac{\lambda}{\Lambda}$ where θ denotes the incident angle, the suffix m denotes the diffraction order, λ denotes the wavelength and Λ denotes the grating period. A guided mode can be excited if the effective index $n_{eff}$ is in the range:

$$\max\{n_{\text{sup}}, n_{\text{sub}}\} \leq |n_{\text{eff}}| < n_{\text{core}}. \tag{1}$$

Out of this range, the propagating modes are not guided. Thus, for the m[th] diffraction order that corresponds to a phase-matched leaky guided mode supported by the resonance structure, the effective refractive index of the waveguide should satisfy to:[22]

$$\max\{n_{\text{sup}}, n_{\text{sub}}\} \leq |n_{\text{sup}} \sin\theta - m\frac{\lambda}{\Lambda}| < n_{\text{core}} \qquad (2)$$

This expression defines parametric regions within which the guided-mode resonances occur. Within the interval designated as the resonance regime, the order *m* can, in accordance with inequality (2), correspond to a guided mode. Another condition that must be respected is the phase matching condition of a slab waveguide, because an electromagnetic wave can be propagating in a waveguide only if the light rays which are totally reflected forms constructive interferences to allow the propagation. In the case of TE-polarization, this phase matching condition can be expressed as[33]:

$$\tan\left(\frac{1}{2}\left(kn_{\text{core}}t\sin\phi - m\pi\right)\right) = \frac{\sqrt{n_{\text{core}}^2 \cos^2\phi - n_{\text{clad}}^2}}{n_{\text{core}}\sin\phi} \qquad (3)$$

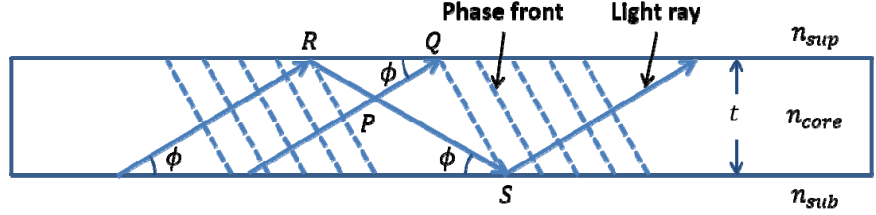

Figure 9: Light rays and their phase fronts in the waveguide.

Fig. 9 illustrates the phase matching condition. Since points P and R or points Q and S are on the same fronts, optical paths PQ and RS should be equal, or their difference should be an integral multiple of $2\pi$.[33] Since the light ray has been incoupled in the waveguide by a diffraction grating, it is possible to combine the grating equation in Eq. 3 and make explicit the thickness dependence of the resonances:

$$t = 2\frac{\frac{m\pi}{2} + \arctan\sqrt{\frac{n_{\text{core}}^2 - n_{\text{clad}}^2}{n_{\text{core}}^2 \cdot \sin^2\left(\arccos\frac{n_{\text{clad}} \cdot \sin\theta + m\frac{\lambda}{\Lambda}}{n_{\text{core}}}\right)} - 1}}{\frac{2\pi}{\lambda} \cdot n_{\text{core}} \cdot \sin\left(\arccos\frac{n_{\text{clad}} \cdot \sin\theta + m\frac{\lambda}{\Lambda}}{n_{\text{core}}}\right)}$$

(4)

In case of TM-polarization, the equation of the resonance regimes are approximated by the following equation:

$$t = 2\frac{\frac{m\pi}{2} + \arctan\frac{n_{\text{core}} \cdot \sqrt{n_{\text{core}}^2 \cos^2\phi - n_{\text{clad}}^2}}{n_{\text{clad}}^2 \sin\left(\arccos\frac{n_{\text{clad}} \cdot \sin\theta + m\frac{\lambda}{\Lambda}}{n_{\text{core}}}\right)}}{\frac{2\pi}{\lambda} \cdot n_{\text{core}} \cdot \sin\left(\arccos\frac{n_{\text{clad}} \cdot \sin\theta + m\frac{\lambda}{\Lambda}}{n_{\text{core}}}\right)}$$

(5)

Eq. 4 and Eq. 5 well describe the behavior of the resonance regimes for a resonant waveguide grating having shallow grooves. For waveguide-gratings having strong corrugation, additional correction must be applied.

In Fig. 10 we report two diagrams based on these two models. The first diagram exhibits the resonance regimes of a RWG with light impinging in normal incidence. We report the resonant regime relative to the first and the second diffracted orders, that can be coupled in the waveguide in the fundamental mode (TE0 or TM0), or higher order modes. In the second diagram, we illustrate the angular dependency of these waveguide gratings. When the incidence angle is tilted at 10° in the superstrate, a separation of the resonance curves for positive and negative diffracted orders appears. For simplicity, resonances generated by second order diffractions are not presented.

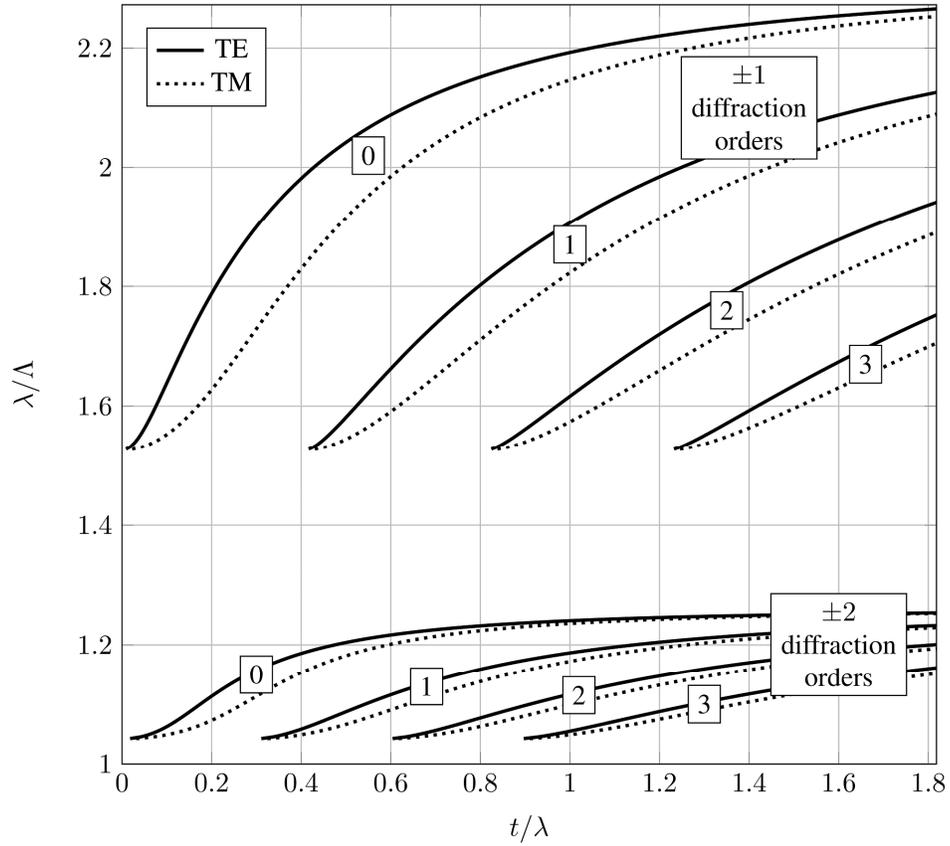

(a) Normal incidence, first and second diffracted orders ($\pm 1, \pm 2$) of the grating are shown.

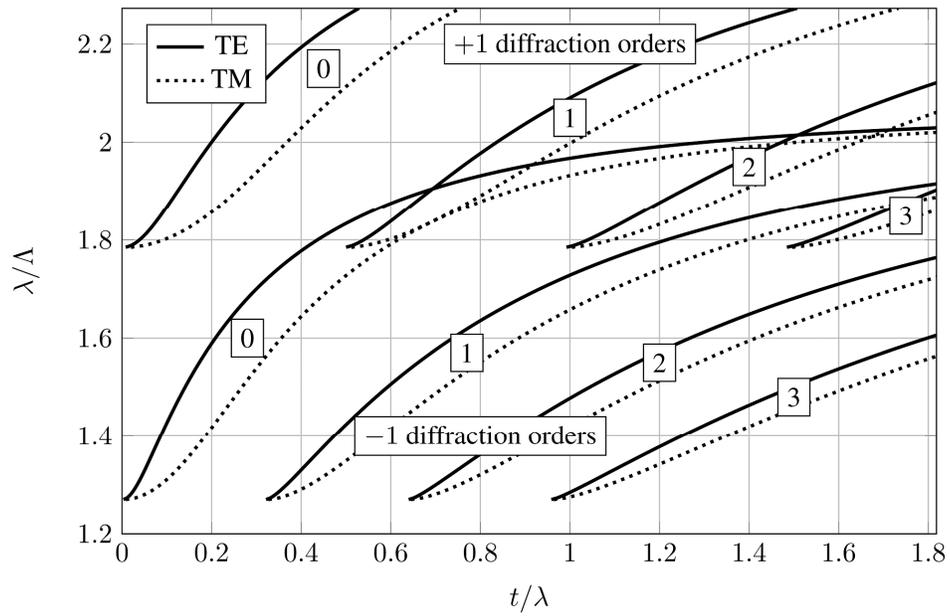

(b) 10° incidence, only the first diffracted orders ($\pm 1$) of the grating are shown.

Figure 10: Resonance regimes of a resonant waveguide grating for different illumination angles.

As can be seen, positive and negative diffracted orders or first and second diffracted orders can be present simultaneously in a well-engineered structure.

### 2.3 New Optical Effects Out Of Specular Angles

It is possible to implement these subwavelength waveguide-gratings to redirect light out of the established Zero-order read-out. Based on multiple resonant waveguide gratings, having different grating vectors, light can be coupled in a waveguide grating in a leaky mode and be out-coupled by another waveguide-grating. Both waveguide gratings can be used simultaneously to incouple a fraction of an incident illumination which will be out-couple by the second waveguide-grating at a designed angle. In Fig. 11 we schematically represents such a light redirection, by resonant diffractive coupling propagation and de-coupling out of the Zero-order read-out (specular reflection):

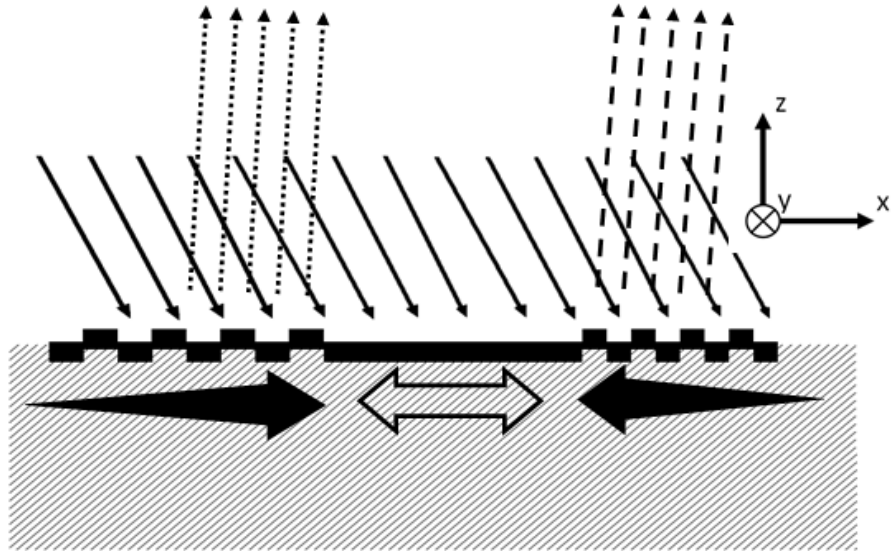

Figure 11: Schematic cross-section of a guided mode resonance device made of two cross-talking RWG[24]

In order to redirect light on a large area, an array of these device can be used, as illustrated in Fig. 12.

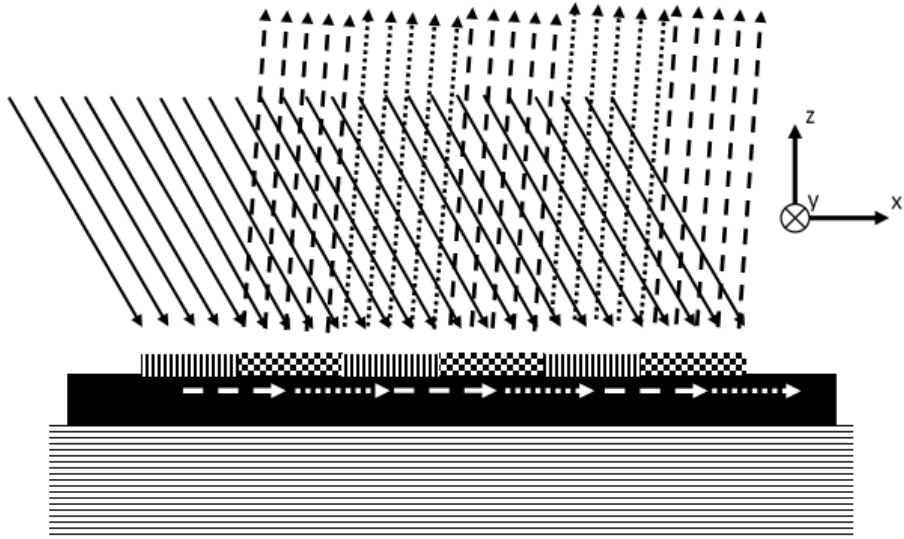

Figure 12: A schematic view of a linear array of a guided mode resonant device made with two elementary structures.[24]

This cross-talking RWG can be used in a variety of applications. In optical security, an interesting implementation is to provide light-redirection for spectral lines engineered with the wanted sharpness. Multi spectral-line and multi polarization devices can be developed as illustrated in Fig. 10.

Using such approaches, we are confident that highly secured DOVIDs can be produced using sub-wavelength waveguide gratings beyond the successful Zero-order layout scheme.

## 3. THICK WAVEGUIDES FOR SECURED DOCUMENTS

### 3.1 From Optical Fibers To Substrates As Multimode Waveguides

Beyond thin resonant waveguide-gratings, security documents are using transparent materials that can be used as optical media. The different technical layers and substrate of passport laminates, banknotes, polycarbonate ID cards and tamper-evident labels could be secured in a new paradigm by transporting light in the bulk of these layers. Previous work aiming at securing documents by using integrated waveguides have been reported in the last half-century. We report here a non-exhaustive summary of the published work known to the authors, mostly in the form of patents and patent applications.

First investigation in the 1970s proposes the integration of optical fibers to secure data cards such as access badges, the fibers being placed in the polymer bulk of the cards. Optical keys were form by the arrangement of the fibers, and their arrangement were read at the card edges[25,26]. A similar idea was applied to credit-cards in 1985 by Photon Devices[27] and to ID documents by Orell Füssli with the integration of fibers between laminated layers[28]. A month before Photon Devices filled its patent application, Arjomari-Prioux, now part of Arjowiggins, described the integration of plastic optical fibers in secured documents such as paper banknotes[29]. For the first time, illumination and light trapping from the top surface of a document is described, although the light-trapping mechanism is not explained.

After this early work, Peter Modler proposed to secure gaming chips and tokens with a similar concept [30]. For the first time, planar waveguides in secured and valued items is proposed, but using injection-molded thermoplastic waveguides thanks to their thicker geometry. Stephen Phillips proposed to couple ambient light in waveguiding layer of documents with chromatic filtering and chromatic angular variability at decoupling, a first concept of waveguiding OVDs[31]. A few years later, the Note Printing of Australia investigated light-guiding in banknotes[32], the Banque De France added a photo-luminescent material in the waveguide[33] and Orell Füssli proposed to add a light emitting diode (LED) to the light-guiding banknote[34].

In the last eight years, the patent activity became more sustained. Among other developments, the Bundesdruckerei proposed to create the waveguide in a security document by printing its core[35], to outcouple light by printing scattering material on the waveguide[36], to print material to modulate spatially the outcoupling light[37] or to incorporate in the substrate light-guiding strands whose both extremities are opened on the top/bottom faces[38]. Orell Füssli Security proposed to couple light in light-guides using its Microperf technology[39] as well as an array of microlenses on top of a patterned coupler[40]. Recently, Hueck Folien proposed to combine light-guiding and see-through security features[41] while Arjowiggins used the waveguide to illuminate a photodiode powered chip[42].

This summary is far from being exhaustive. Patent applications are disclosing various technical possibilities and cannot be fully summarized in a single idea. It is worth noting that the rise of polymer banknotes and of polycarbonate ID documents did not generate a large patent activity about waveguides in security documents. On the opposite, the patent activity acceleration matches well the fast smartphone adoption started in 2007, providing both a wide light source and a high definition camera to the vast majority of the population.

### 3.2 Effective Optical Couplers As Key Enablers

In the last three years, CSEM developed and reported new classes of diffractive optical couplers. These couplers are based on subwavelength gratings, especially fabrication-friendly binary and sinusoidal gratings. These gratings are designed to be compatible with existing roll-to-roll nano-imprinting facilities. We coat them with a subwavelength coating made of a HRI dielectric or a metal in order to reach high diffraction efficiencies as well as to fully embed them. The coatings are similar to the standard deposition made in evaporators for the metallization of holograms or the coating of transparent surface holograms.

In a first development, we optimized the coating of undemanding subwalength gratings to reach high diffraction efficiencies over broad wavelength ranges[43]. The period is chosen so that the incident light which is diffracted at the first order propagates in the layer below the grating at flat angles under Total Internal Reflection (TIR) with minor refractive index contrasts. In detail, the structures can be optimized to diffract the incident light over a selected spectral region, for unpolarized or polarized light and over a broad solid angle of incidence. These extra flat nanostructures are designed to be replicated and coated at high speed as well as embedded with lacquers, on the opposite to previously reported highly efficient slanted grating coupler[44, 45]. This allows both a high mechanical stability over the coupler area to the laminate and prevent their direct replication. We present in Fig. 13 the diffraction efficiency of a sinus profile grating coated with Zinc Sulfide.

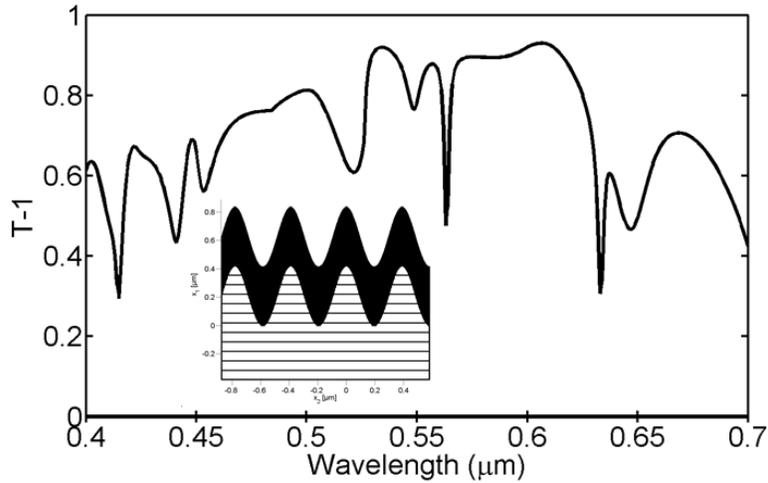

Figure 13: First-order transmission diffraction efficiency for unpolarized light impeding at 17° on a sub-wavelength sinusoidal-profile grating of 390nm period coated with a HRI dielectric layer. Because of the small period of the grating, the light diffracted in transmission is trapped in the layer immediately below it in TIR.

The major inconvenient of these subwavelength gratings and their coating is their limited efficiency close to normal incidence. Because of their symmetry, the transmitted positive and negative first order of diffraction have similar efficiency which is therefore limited to below 50% of the incident illumination. In order to achieve higher coupling efficiencies at normal incidence, without requiring complex mastering and replication processes, we proposed to create the asymmetry with the subwavelength coating. On simple binary or sinusoidal gratings, we reported high diffraction efficiencies by evaporating a thin metal[46] or HRI dielectric[46] obliquely, as presented respectively in Fig. 14 and Fig. 15.

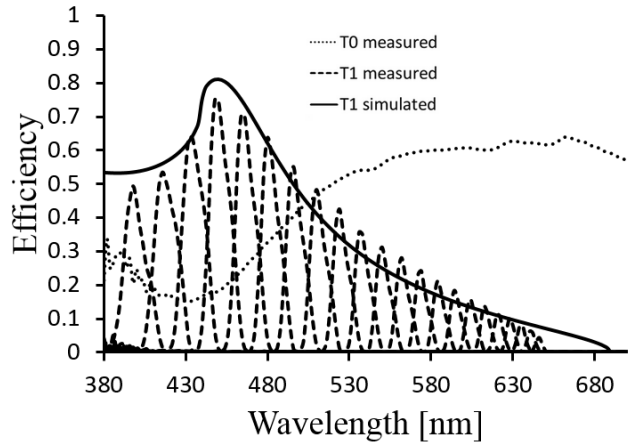

Figure 14: Measured first-order transmission diffraction efficiency versus numerically computed one for an embedded 440nm period grating coated with an aluminum coating (Al) for a TE polarized light incident normal to the grating[46]

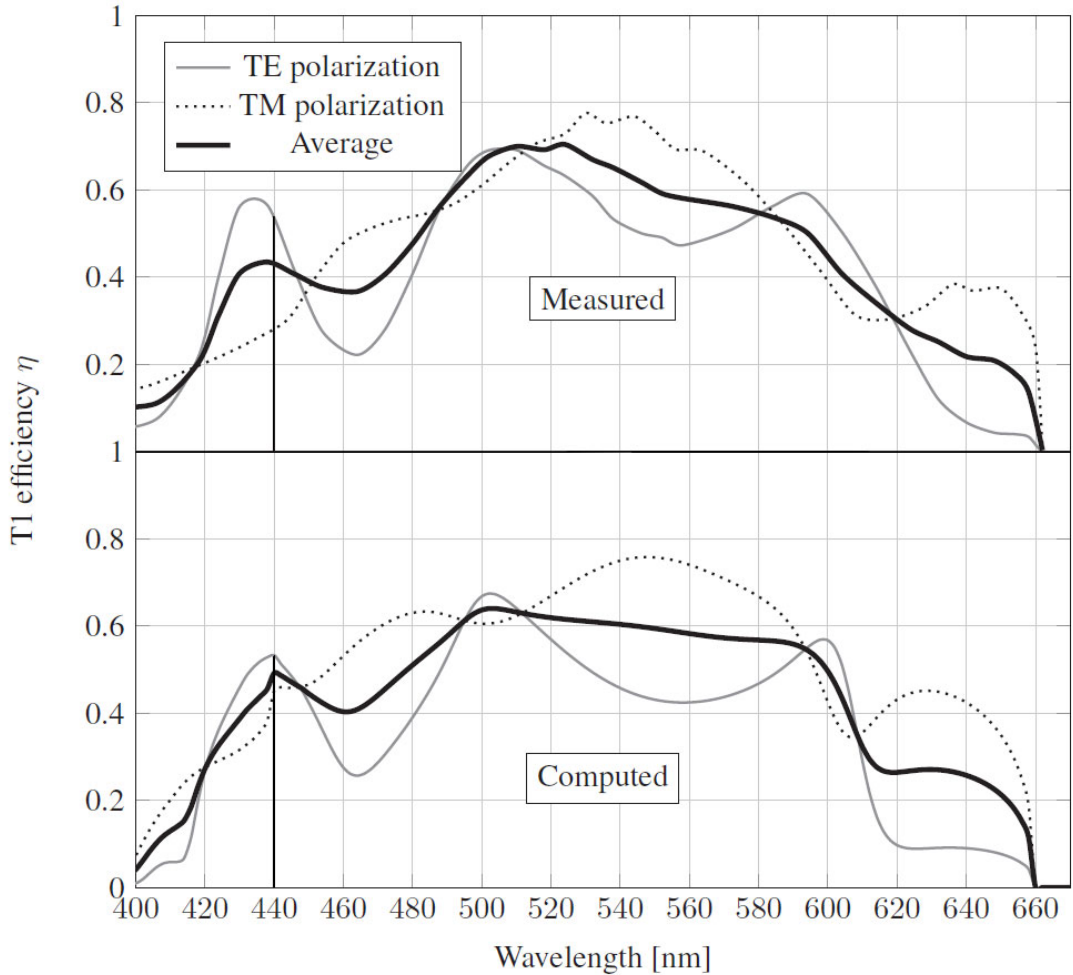

Figure 15: Measured first-order transmission diffraction efficiency versus numerically computed one for an embedded 440nm period symmetric grating coated with a Zinc Sulfide coating (ZnS) for a light incidence normal to the grating[46].

These subwavelength grating couplers can be produced with existing advanced holograms fabrication facilities, be totally embedded in a layer stack of laminate and provide efficient, broadband and broad-angle light in or out-coupling. We believed such designs can boost the use of the transparent layer of security documents as waveguides. Most people are today equipped with a white LED and high-definition camera in their smartphone. These are already widely used for document security communication, education and authentication. We believe their capabilities can be used more thoroughly.

### 3.3 Outlook, The Example Of Tamper-Evident Seals

Security document are currently secured with many different technologies. Most experts acknowledge that overt features are the most challenged as counterfeit or forged documents are in most cases not breaking covert security features. Often, simple imitations are used as counterfeit as they are sufficient to access a fraction of the counterfeited market as a distribution channel.

The visual or smartphone-based control of secured tamper-evident seals, most of which are tamper-evident sticker labels, can often be circumvented by a clean razorblade cut breaking-in, followed by an edge to edge cementing of the label. This easy method can enable the resell of counterfeited packages to be sold as genuine and challenge naked-eye scrutiny as well as smartphone based algorithm responsible for authentication of the labels. In such example, implementing a light-guiding structure between the two ends of the seals, attached to the two part of the object to be secured can demonstrate a break-in. Attempts at seal opening, such as a razorblade cut, would create a loss of light-guiding even for micro-scale breaks. Microscopic cuts and defects can be made well visible at a macroscopic scale. Cementing edge to edge a cut seal containing a light-guide will not be sufficient to reestablish a clean optical contact. Additionally various features can provide extra security to prevent optical cementing attempts such as using multiple independent waveguide or making visible unwanted light propagating in a cladding.

On the opposite to a large number of security features, lightguide in documents can provide, in addition to a secured authentication, an evidence of integrity for the bulk of the secured document - both in its lateral dimension and in its volume. Ubiquitous smartphones and emerging wearable electronics are perfectly suited to bring light in a secured document and to analyze its optical output. The interactivity of such control can catch attention as well as being fully guided and authentified by an application. We believe that efficient, easy to fabricate and robust optical couplers based on subwavelength gratings can help giving security documents the extra functionality of light-guiding.

## ACKNOWLEDGMENTS

The authors would like to express their deepest gratitude to the pioneers of subwavelength gratings, especially Karl Knop.